\documentclass[12pt]{iopart}
\usepackage{iopams}
\usepackage{graphicx}
\usepackage{color}
\usepackage[colorlinks=true,linkcolor=blue,pagecolor=blue,filecolor=blue,menucolor=blue,urlcolor=blue,citecolor=blue,anchorcolor=blue]{hyperref}
\usepackage{orcidlink}
\begin{document}

\title[]{Quantum-state engineering in cavity magnomechanics formed by two-dimensional magnetic materials}
\author{Chun-Jie Yang$^1$\orcidlink{0000-0003-2137-6958}, QingJun Tong$^{2,\ast}$\orcidlink{0000-0002-3500-1228}, and Jun-Hong An$^{3,\ast}$\orcidlink{0000-0002-3475-0729}}
\address{$^1$ School of Physics, Henan Normal University, Xinxiang 453007, China}
\address{$^2$ School of Physical Science and Electronics, Hunan University, Changsha 410082, China}
\address{$^3$ Key Laboratory of Quantum Theory and Applications of MoE, Lanzhou University, Lanzhou 730000, China}
\ead{tongqj@hnu.edu.cn and anjhong@lzu.edu.cn}
\vspace{10pt}

\begin{abstract}
Cavity magnomechanics has become an ideal platform to explore macroscopic quantum effects. Bringing together magnons, phonons, and photons in a system, it opens many opportunities for quantum technologies. It was conventionally realized by an yttrium iron garnet, which exhibits a parametric magnon-phonon coupling $\hat{m}^\dag\hat{m}(\hat{b}^\dag+\hat{b})$, with $\hat{m}$ and $\hat{b}$ being the magnon and phonon modes. Inspired by the recent realization of two-dimensional (2D) magnets, we propose a cavity magnomechanical system using a 2D magnetic material with both optical and magnetic drivings. It features the coexisting photon-phonon radiation-pressure coupling and quadratic magnon-phonon coupling $\hat{m}^\dag\hat{m}(\hat{b}^\dag+\hat{b})^2$ induced by the magnetostrictive interaction. A stable squeezing of the phonon and bi- and tri-partite entanglements among the three modes are generated in the regimes with a suppressed phonon number. Compared with previous schemes, ours does not require any extra nonlinear interaction and reservoir engineering and is robust against the thermal fluctuation. Enriching the realization of cavity magnomechanics, our system exhibits its superiority in quantum-state engineering due to the versatile interactions enabled by its 2D feature.
\end{abstract}

%

%
\maketitle
%

\section{Introduction}\label{introduction}
Hybrid quantum systems with multiple degrees of freedom are widely used in exploring fundamental physics and building novel functional quantum devices \cite{RevModPhys.85.623,Doi:10.1073/pnas.1419326112,PhysRevLett.107.220501,Clerk2020}. The heart of these applications is the designing of coherent couplings between different degrees of freedom in these hybrid systems. Cavity magnomechanics has emerged as an ideal platform to study the coherent interactions between photons, phonons, and magnons \cite{cavitymagnonopto,Lachance_Quirion_2019,ZARERAMESHTI20221,doi:10.1063/5.0046202,YUAN20221,PhysRevLett.124.093602,PhysRevLett.125.117701,PhysRevLett.125.147201,PhysRevLett.128.013602,Pirro2021,PhysRevApplied.17.034024,PhysRevX.11.031053,PhysRevLett.124.213604,PhysRevApplied.13.014053,ZHANG2023100044,doi:10.1142/11820}. Fascinating applications have been envisioned, such as low-temperature thermometer \cite{PhysRevApplied.13.064001}, quantum memory for photonic quantum information \cite{Sarma_2021}, and building block for long-distance quantum network \cite{PRXQuantum.2.040344}.

Conventional cavity magnomechanics is realized by yttrium iron garnets (YIGs), where the phonon linearly couples to the magnon as a result of its isotropic magnetostrictive interaction. However, the emergence of many quantum phenomena, e.g., the mechanical bistability \cite{PhysRevLett.129.123601}, squeezing generation \cite{PhysRevResearch.3.023126}, and nonreciprocal magnetic transmission \cite{PhysRevApplied.12.034001}, require carefully engineered nonlinearity. Thus, current cavity magnomechanical systems dramatically resort to complex ``self-Kerr'' nonlinearity \cite{PhysRevB.94.224410} or the squeezed reservoir injection \cite{PhysRevA.99.021801}. Developing new platforms to realize cavity magnomechanics with versatile magnon-phonon couplings is greatly desired for their practical applications in quantum technologies \cite{YUAN20221}. Recently, atomically thin two-dimensional (2D) materials have become an exciting platform for exploring low-dimensional physics and functional devices \cite{Doi:10.1126/science.aac9439,Liu2019}. With the advent of 2D magnets \cite{Wang2016,Du2016,Lee2016,Gong2017,Huang2017,Burch2018,doi:10.1126/science.aav4450,Mak2019,Huang2020}, it is now possible to add the magnon degree of freedom in these atomically thin mechanical systems \cite{Jiang2020,Siskins2020,Cenker2021}. Building such a 2D hybrid optical, magnonic, and mechanical system has immediate advantages over the existing cavity magnomechanical systems based on the YIG. First, a 2D magnet has an out-of-plane flexural phonon mode that may exhibit possible high-order coupling to the magnon due to its highly anisotropic magnetostrictive interactions \cite{PhysRevLett.100.076801,PhysRevB.89.184413,CHENG20081}, which is key to the quantum-state engineering based on cavity magnomechanics. Second, a mechanical oscillator made of a 2D magnet is sensitive to external forces due to its low mass \cite{Doi:10.1126/science.1136836,Chen2013,Lee2013,Morell2016,Morell2019}, which induces a photon-phonon radiation pressure absent in the existing cavity magnomechanics \cite{doi:10.1126/science.1156032,RevModPhys.86.1391}. Therefore, 2D magnetic materials may open another avenue to realize cavity magnomechanics.

\begin{figure}[tbp]
\centering
\includegraphics[width=15cm]{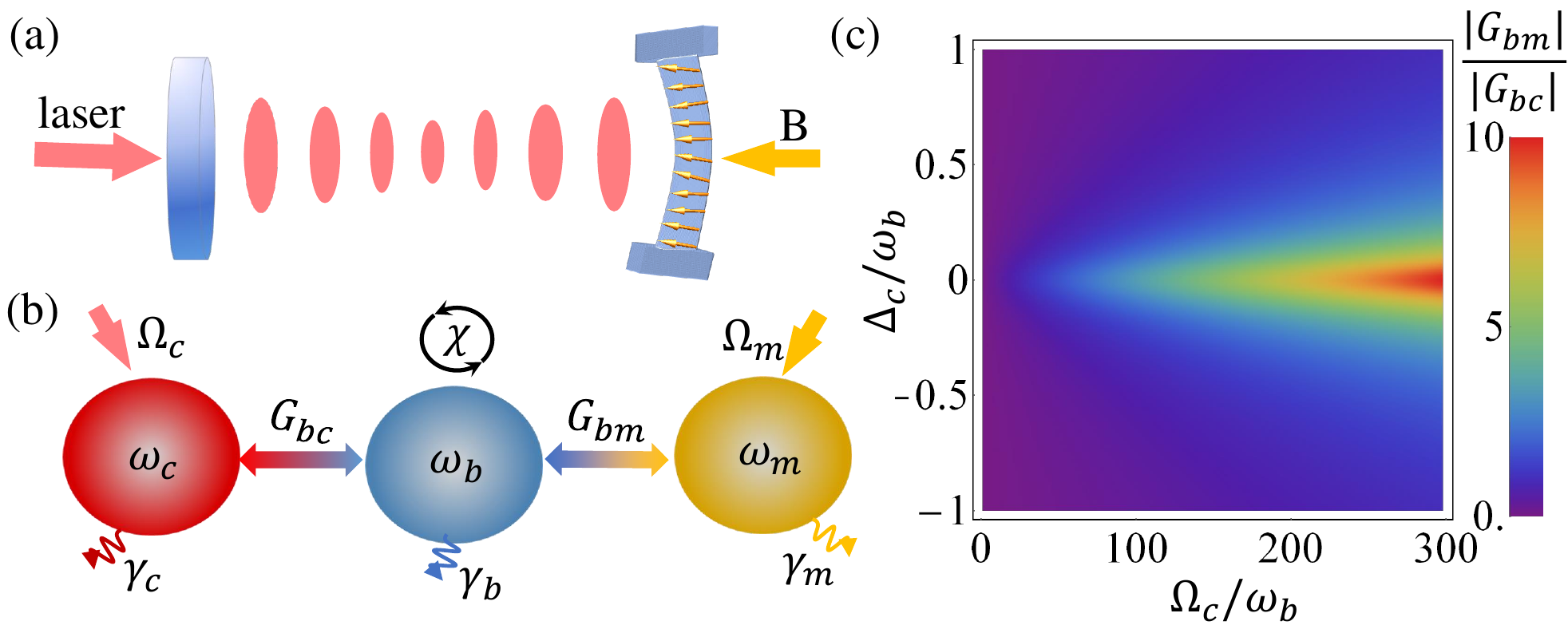}\\
\caption{(a) Schematics of a cavity magnomechanical system: An optically driven cavity interacted with a magnetic membrane under a magnetic driving. (b) Interactions among the photons, phonons, and magnons with frequencies $\omega_o$ and damping rates $\gamma_o$ ($o\in\{c,b,m\}$).
A parametric amplification to the phonon mode with strength $\chi$ is induced. The photon-phonon coupling $G_{bc}$ and the magnon-phonon coupling $G_{bm}$ are enhanced by the optical and magnetic drivings.(c) $|G_{bm}/G_{bc}|$ as a function of $\Delta_c$ and $\Omega_c$. The parameters are from Fig. \ref{squzphon}(b).}
\label{fig1}
\end{figure}
We propose such a cavity magnomechanical system using a 2D magnetic material with both optical and magnetic drivings. A quantized description reveals that this hybrid system has a combined parametric optomechanical and quadratic magnomechanical interactions. The unique photon-phonon-magnon interaction endows our system with the distinguished role in quantum-state engineering. We find that a stable phonon squeezing, and bi- and tri-partite entanglement among the three modes are generated in the regimes with a suppressed phonon number. Steming from the unique magnomechanical coupling in 2D magnets, the generation of these quantum effects requires neither the ``self-Kerr'' nonlinearity nor the squeezed-reservoir engineering. More importantly, due to the accompanied suppressed phonon number, our scheme is robust against thermal noise, which exhibits a superiority over the conventional ones.

\section{Model Hamiltonian}\label{sys}
We consider a hybrid system of cavity optomechanics and 2D magnetic material. The system consists of an optically driven cavity interacted with a 2D magnetic membrane [see Fig. \ref{fig1}(a)]. Its spin interactions induce a collective wave, which couples to the mechanical deformation of the magnet. The induced magnetoelastic energy is $\mathcal{E}=\frac{\sigma }{\mathcal{M}^{2}_S}\int dV\sum_{\alpha\beta }B_{\alpha \beta }\mathcal{M}_{\alpha }\mathcal{M}_{\beta }U_{\alpha \beta }(\mathbf{r})$, where $\sigma=N/V$ is the number density of the magnetic atoms with $N$ and $V$ being the number of magnetic atoms and the volume of the magnet, $\mathcal{M}_S$ is the saturation magnetization, $\mathcal{M}_{\alpha }$ $(\alpha=x,y,z)$ are the local magnetization, $B_{\alpha \beta }$ are magnetoelastic coupling constants, and $U_{\alpha \beta}(\mathbf{r})=[\partial_{\beta} U_{\alpha }(\mathbf{r})+\partial_{\alpha} U_{\beta }(\mathbf{r})+\sum_{\gamma }\partial_{\alpha} U_{\gamma }(\mathbf{r})\partial_{\beta} U_{\gamma }(\mathbf{r})]/2$ are the strain tensors for the lattice displacement $U(\mathbf{r})$ \cite{RevModPhys.21.541,PhysRev.110.836}. The second-order term in $U_{\alpha \beta}(\mathbf{r})$ is negligible in 3D systems (like YIG), which leads to a radiation-pressure-like linear magnon-phonon coupling \cite{cavitymagnonopto}. However, the 2D nature of a magnetic membrane creates a unique flexural phonon mode, which makes the second-order term important \cite{PhysRevLett.129.123601}. Assuming $U_{xx}(\mathbf{r})=U_{yy}(\mathbf{r})$ and Fourier transforming $U_{\alpha \beta}(\mathbf{r})$, the energy regarding flexural magnon-phonon coupling reads (see \ref{mag-pho-int})
\begin{eqnarray}\label{mageef}
\mathcal{E} &=&\frac{\sigma B_{1}}{2\mathcal{M}_{S}^{2}}(\mathcal{M}_{S}^{2}-%
\mathcal{M}_{z}^{2})k_{x}^{2}|\tilde{U}_{z}(k)|^{2}  \label{mageef} \nonumber\\
&&+\frac{\sigma B_{2}}{2\mathcal{M}_{S}^{2}}\Big[(\mathcal{M}_{x}\mathcal{M}%
_{y}+\mathcal{M}_{y}\mathcal{M}_{x})k_{x}k_{y}|\tilde{U}_{z}(k)|^{2},
\end{eqnarray}
where $B_{1}=B_{\alpha \alpha}$ and $B_{2}=B_{\alpha \beta }$ ($\alpha\neq\beta$). Introducing the phonon $\tilde{U}_{z}(k)={[\hbar/(2m_{0}\omega _{b})]}^{1/2}(\hat{b}+ \hat{b}^\dag)$ and the magnon $\mathcal{M}_{\alpha}= 2\mu_{B}\sigma \hat{S}_{\alpha}$, with $\hat{S}_{x} =2^{-1/2}S^{1/2}(\hat{m}+\hat{m}^{\dag})$, $\hat{S}_{y}= -i2^{-1/2}S^{1/2}(\hat{m}-\hat{m}^{\dag })$ and $\hat{S}_{z} =S-\hat{m}^{\dag }\hat{m}$, the first term regarding to the dispersive interaction reduces to
\begin{eqnarray}\label{hbm}
\hat{H}_{bm}/\hbar&=&g_{bm}\hat{m}^{\dag }\hat{m}(\hat{b}^{\dag }+\hat{b})^{2},
\end{eqnarray}
with the coupling strength $g_{bm}=\frac{B_{1}}{4S}\frac{ k_{x}^{2}}{2m_{0}\omega _{b}}$, where $m_0$ is the ion mass, $\omega_{b}$ and $k_x$ are the resonance frequency and wave vector of the mechanical mode. The second term of Eq. (\ref{mageef}) is negligible when the resonance frequency of the phonon is much smaller than that of the magnon \cite{cavitymagnonopto}. The quadratic magnon-phonon coupling in Eq. (\ref{hbm}) is absent in 3D magnets, which is associated with the 2D nature of the membrane. The photon also exerts a radiation pressure on the magnetic membrane, which triggers the photon-phonon coupling $\hat{H}_{bc}/\hbar=g_{bc}\hat{c}^{\dag }\hat{c}(\hat{b}^{\dag }+\hat{b})$ (see \ref{coupling-strength}). Our system not only gives a novel realization of the rapidly developing cavity magnomechanics, but also generalizes the conventional systems into a quadratic magnon-phonon coupling regime. We note that this non-linear interaction may lead to parametric instability or chaotic dynamics in evolution of the coupled system \cite{PhysRevLett.96.103901,PhysRevLett.98.167203}, and can also be used to engineer non-Gaussian states of the magnons and phonons \cite{PRXQuantum.2.030204}.

In the rotating frame with $\hat{H}_0=\hbar\omega_{dc}\hat{c}^\dag\hat{c}+\hbar\omega_{dm}\hat{m}^\dag\hat{m}$, the total Hamiltonian in the presence of both optical and magnetic drivings reads
\begin{eqnarray}
\hat{H}/\hbar &=&\Delta_{c}\hat{c}^{\dag }\hat{c}+\Delta _{m}\hat{m}^{\dag }\hat{m}+\omega _{b}\hat{b}^{\dag }\hat{b}+g_{bc}\hat{c}^{\dag }\hat{c}(\hat{b}+\hat{b}^{\dag }) \nonumber\\
 &&+g_{bm}\hat{m}^{\dag }\hat{m}(\hat{b}^{\dag }+\hat{b})^{2}+(\Omega _{c}\hat{c}^{\dag }+\Omega _{m}\hat{m}^{\dag }+\textrm{H.c.}).
\end{eqnarray}
Here $\hat{c}$ is the annihilation operator of the cavity with frequency $\omega_{c}$, $g_{bc}$ is the photon-phonon coupling strength, and $\Delta_{c/m}=\omega_{c/m}-\omega_{dc/dm}$ are the photon and magnon detunings to their driving frequencies $\omega_{dc/dm}$. The Rabi frequencies of the driving fields on the cavity and magnet are $\Omega _{m}=\gamma B_{0}\sqrt{2NS}/4$ \cite{PhysRevLett.121.203601} and $\Omega _{c}=\sqrt{P\gamma _{c}/\hbar \omega _{dc}}$ \cite{PhysRevLett.99.093901}, where $\gamma/2\pi=28$ GHz/T is the gyromagnetic ratio, $B_{0}$ is the amplitude of the drive magnetic field, $P$ is the input laser power, and $\gamma_c$ is the damping rate of the cavity mode. We have eliminated the direct magnon-photon interaction in the model, as the coupling strength is rather weak due to the significantly reduced number of spins in 2D system. The dynamics is governed by the master equation
\begin{eqnarray}\label{maseq}
\dot{W}(t) &=&\frac{i}{\hbar }[W(t),\hat{H}] \nonumber\\
&&+[\gamma _{c}\check{\mathcal{L}}_{\hat{c}}+\gamma _{m}\check{\mathcal{L}}_{
\hat{m}}+\gamma _{b}\bar{n}_{0}\check{\mathcal{L}}_{\hat{b}^{\dag }}+\gamma
_{b}(1+\bar{n}_{0})\check{\mathcal{L}}_{\hat{b}}]W(t),
\end{eqnarray}
where $W(t)$ is the density matrix of the three-mode system, $\check{\mathcal{L}}_{\hat{o}}\cdot=2\hat{o}\cdot\hat{o}^\dag-\{\hat{o}^\dag\hat{o},\cdot\}$ is the Lindblad superoperator, $\gamma_o$ are the damping rates of the three bosonic modes, and $\bar{n}_0=[\exp(\hbar\omega_b/k_BT)-1]^{-1}$ is the mean thermal excitation number of the environment felt by the phonon.

The strong optical and magnetic drivings make the steady-state occupations of the three modes have large amplitudes. This allow us to linearize the Hamiltonian on one hand and enhance both of the magnon-phonon and photon-phonon couplings on the other. From Eq. (\ref{maseq}), the steady-state values of the three modes under the condition $\gamma_b\ll\gamma_m$ are $\bar{c}=\frac{\Omega _{c}}{i\gamma _{c}-\Delta _{c}}$, $\bar{b}\approx-\frac{g_{bc}|\bar{c}|^{2}}{\omega_{b}+4\chi}$, and $\bar{m}=\frac{\Omega _{m}}{i\gamma _{m}-\Delta _{m}}$, with $\chi=g_{bm}|\bar{m}|^{2}$ and $\bar{o}=\textrm{Tr}[W(\infty)\hat{o}]$. Rewriting $\hat{o}=\bar{o}+ \hat{\tilde o}$ and neglecting the fluctuation operators up to the second order, we obtain
\begin{eqnarray}
\hat{H}/\hbar &=&\Delta^{\prime}_{c}\hat{c}^{\dag }\hat{c}+\Delta^{\prime}_{m}\hat{m}^{\dag
}\hat{m}+\omega^{\prime}_{b}\hat{b}^{\dag }\hat{b}+(G_{bc}^{\ast }\hat{c}+G_{bc}\hat{c}^{\dag })(\hat{b}+\hat{b}^{\dag })  \nonumber\\
&&+(G_{bm}^{\ast }\hat{m}+G_{bm}\hat{m}^{\dag })(\hat{b}^{\dag }+\hat{b})+\chi (\hat{b}^{\dag 2}+\hat{b}^{2}),   \label{liner-H1}
\end{eqnarray}
where $\Delta^{\prime}_{c}=\Delta_{c}+2 g_{bc}\bar{b}$, $\Delta^{\prime}_{m}=\Delta_{m}+4 g_{bm}\bar{b}^2$, $\omega^{\prime}_{b}=\omega_{b}+2\chi$, $G_{bc}=g_{bc}\bar{c}$, and $G_{bm}=4g_{bm}\bar{m}\bar{b}$. The tilde of $\hat{\tilde{o}}$ in Eq. (\ref{liner-H1}) has been omitted for brevity. Both of $G_{bc}$ and $G_{bm}$ are renormalized by the steady-state values of the three bosonic modes.

Our system possesses the unique photon-phonon-magnon interactions [see Fig. \ref{fig1}(b)]. First, the magnetostrictive interaction favored by the optical and magnetic drivings induces an effective bilinear magnon-phonon interaction in the fifth term of Eq. (\ref{liner-H1}), which resembles the photon-phonon interaction in the fourth term. Second, the magnon-phonon coupling is jointly enhanced by the magnetic and optical drivings. It is different from the conventional magnomechanics, where the magnomechanical coupling is renormalized only by the magnetic driving \cite{cavitymagnonopto}. This can be properly used to achieve the strong magnon-phonon coupling. Figure \ref{fig1}(c) exhibits that, vanishing in the absence of the optical driving, $|G_{bm}|$ reaches its maximum at the optically resonant driving point $\Delta_{c}=0$. A ten-time's gain of $|G_{bm}|$ over $|G_{bc}|$ is obtained by increasing $\Omega_c$ at $\Delta_{c}=0$. Third, the magnetostrictive interaction and the magnetic driving create a parametric amplification to the phonon mode in the last term of Eq. (\ref{liner-H1}), which is absent in the conventional magnomechanics \cite{cavitymagnonopto}. All of these features endow our system with the superiority to realize quantum-state engineering.

\section{Phonon squeezing}\label{squ}
We first explore the quantum effect of the steady-state phonon. To gain some insight, we analytically derive the steady-state phonon number. Via adiabatically eliminating the photon and magnon modes from Eq. (\ref{maseq}) under $\gamma _{m},\gamma _{c}\gg\gamma _{b}$ \cite{PhysRevLett.121.203601}, we obtain a reduced master equation of the phonon \cite{PhysRevLett.103.063005,PhysRevA.92.062311} (see \ref{red-mast})

\begin{eqnarray}\label{re-mast}
\dot{\rho}(t) &=&i[\rho (t),\tilde{\omega}_{b}\hat{b}^{\dag }\hat{b}+\chi (\hat{b}^{\dag2 }+\hat{b}^2)]\nonumber\\
&&+[(\zeta _{+}+\zeta _{-}^{\ast })\hat{b}\rho (t)\hat{b}-\zeta _{-}^{\ast }\hat{b}^2\rho (t)  -\zeta _{+}\rho (t)\hat{b}^2+\textrm{H.c.}]\nonumber\\
&&+\{[\gamma_{b}(1+\bar{n}_{0})+\zeta_-^{r}]\check{\mathcal{L}}_{\hat{b}}+(\gamma_b\bar{n}_{0}+\zeta_+^{r})\check{\mathcal{L}}_{\hat{b}^{\dag}}\}\rho(t),
\end{eqnarray}
where $\tilde{\omega}_{b}=\omega^{\prime}_{b}-\sum_{x=c,m}(\zeta^{i}_{+x}+\zeta^{i}_{-x})$, $\rho(t)=\textrm{Tr}_{c,m}[W(t)]$, and $\zeta _{\pm}=\sum_{x=c,m}\zeta _{\pm x}$, with $\zeta _{\pm x}=|G_{bx}|^{2}[\frac{e^{-r}\cosh r}{\gamma_{x}-i(\Delta _{x}^{\prime }\pm \bar{\omega})}+\frac{e^{-r}\sinh
r}{\gamma _{x}-i(\Delta _{x}^{\prime }\mp \bar\omega)}]$, $r=\textrm{arccoth}({\omega^{\prime}_{b}\over2\chi})/2$, and $\bar{\omega}={[\omega_b(\omega_b+4\chi)]}^{1/2}$. The superscripts ``$r$'' and ``$i$'' denote the real and imaginary parts. It is interesting to see that the photon and magnon, as two ``contact environments'', can not only induce the thermal-like dissipation to the phonon in the terms $\zeta^r_\pm$ \cite{PhysRevLett.99.093901}, but also induce the incoherent squeezing in the second line \cite{PhysRevLett.127.083602}. Assisted by this incoherent squeezing and the coherent parametric amplification, a stable squeezing of the phonon is expected to be generated.

\begin{figure}[tbp]
	\centering
	\includegraphics[width=15cm]{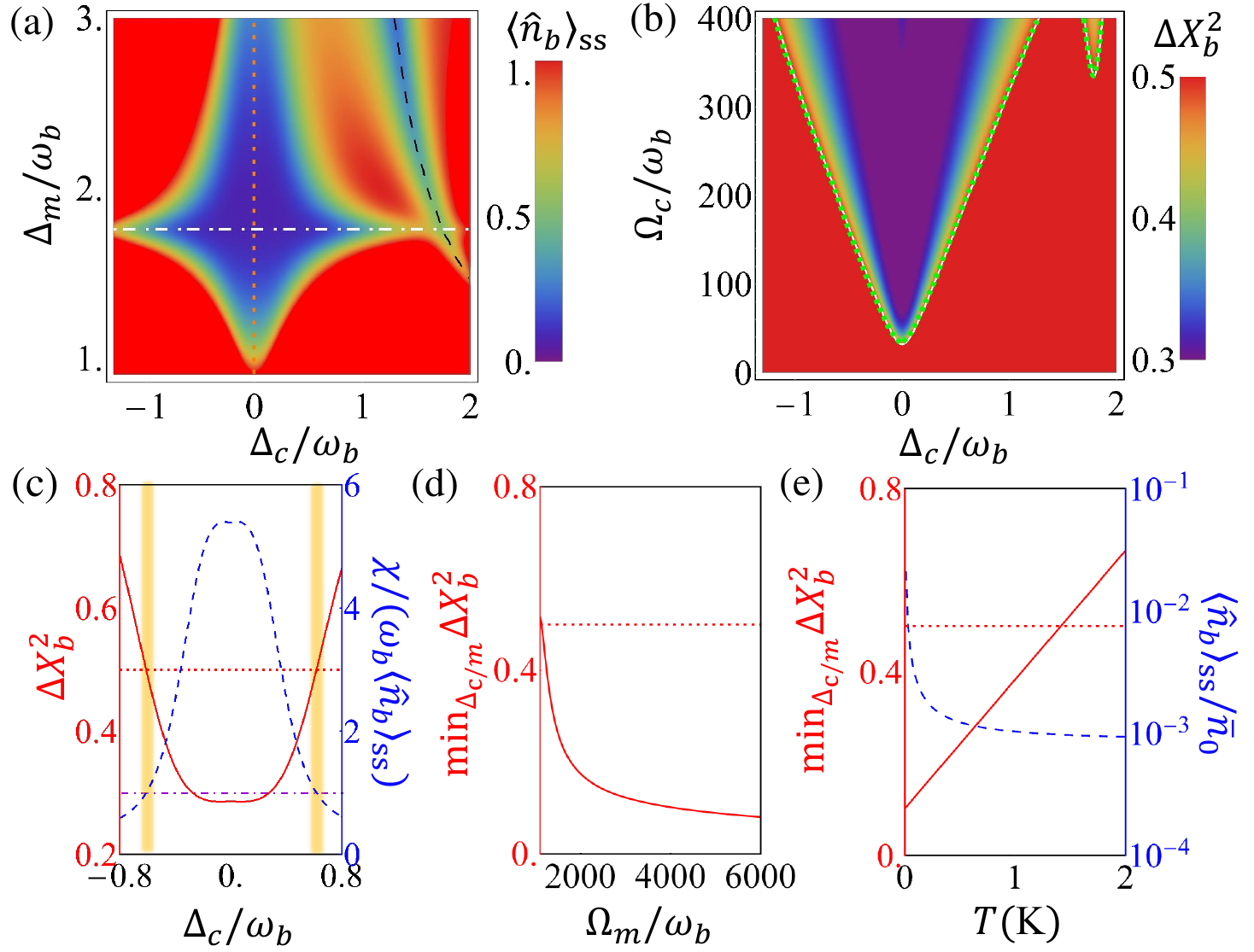}
	\caption{(a) Steady-state phonon number $\langle \hat{n}_{b}\rangle_\textrm{ss}$ as a function of $\Delta_m$ and $\Delta_c$ when $\bar{n}_{0}=1$ and $(\Omega_m,\Omega_c)=(400,300)\omega_b$. The conditions of $\Delta_{c}=0$, $\Delta_{m}'=\bar\omega$, and $\Delta^{\prime}_{c}=\bar\omega$ are depicted by the orange dotted, white dot-dashed, and black dashed lines, respectively. (b) Fluctuation $\Delta X_{b}^{2}$ as a function of $\Omega_c$ and $\Delta_c$ when $\Delta_m=1.69\omega_b$. The white line shows $\Delta X_{b}^{2}=1/2$, which matches with the squeezing boundary $\chi/(\omega_b\langle \hat{n}_b\rangle_\textrm{ss})=1$ (green dots). (c) $\Delta X_{b}^{2}$ (red solid line) and $\chi/(\omega_b\langle \hat{n}_b\rangle_\textrm{ss})$ (blue dashed line) when $\Omega_c=200\omega_b$. $\Delta X_{b}^{2}<1/2$ whenever $\chi/(\omega_b\langle \hat{n}_b\rangle_\textrm{ss})>1$. (d) $\min_{\Delta_{c/m}}\Delta X_{b}^{2}$ as a function of $\Omega_m$ when $\Omega_c=10^3\omega_b$. (e) Temperature dependence of $\min_{\Delta_{c/m}}\Delta X_{b}^{2}$ (red solid line) and $\langle \hat{n}_{b}\rangle_\textrm{ss}/\bar{n}_{0}$ (blue dashed line) when $\Omega_m=4000\omega_b$. Other parameters are $(\gamma_b,\gamma_c,\gamma_m)=(10^{-5},0.1,0.1)\omega_b$ and $g_{bc}=g_{bm}=10^{-5}\omega_b$ with $\omega_b=25$ MHz and $\omega_m=2\pi\times120$ GHz \cite{Jiang2020,PhysRevLett.124.017201}.} \label{squzphon}
\end{figure}

Squeezing is characterized by the reduced fluctuation of the quadrature operator $\hat{X}_{b}^{\theta}=\cos\theta \hat{X}_{b}+\sin\theta \hat{Y}_{b}$, with $\hat{X}_b=(\hat{b}+\hat{b}^\dag)/\sqrt{2}$ and $\hat{Y}_b=(\hat{b}-\hat{b}^\dag)/\sqrt{2}i$, in certain $\theta\in[0,2\pi)$ below the standard quantum limit, i.e., $\min_{\theta}\Delta X_{b}^{\theta 2}<1/2$. It is derived from Eq. (\ref{re-mast}) that
\begin{eqnarray}
\langle\hat{n}_b\rangle_\textrm{ss}&=&{\gamma_b\bar{n}_0+\zeta_+^{r}\over\gamma_b+\zeta_-^{r}-\zeta_+^{r}}+\frac{\textrm{Re}[(\zeta _{-}^{\ast }-\zeta
_{+}+2i\chi )\langle \hat{b}^{2}\rangle_\textrm{ss} ]}{\gamma _{b}+\zeta _{-}^{r}-\zeta
_{+}^{r}}, \label{phmb}\\
\langle \hat{b}^{2}\rangle_\textrm{ss} &=&\frac{(\zeta _{-}-\zeta _{+}^{\ast }-2i\chi )\langle
\hat{b}^{\dag }\hat{b}\rangle_\textrm{ss} -(i\chi +\zeta _{+}^{\ast })}{\gamma _{b}+i\tilde{\omega}_{b}+\zeta _{-}^{r}-\zeta _{+}^{r}},
\end{eqnarray}
with $\hat{n}_b=\hat{b}^\dag\hat{b}$ and $\langle\bullet\rangle_\textrm{ss}=\textrm{Tr}[\rho(\infty)\hat{\bullet}]$. The first term of Eq. (\ref{phmb}) is contributed from the thermal-like dissipation in the last line of Eq. (\ref{re-mast}) \cite{PhysRevLett.99.093901,PhysRevA.77.033804}. The second term is from the incoherent squeezing in the second line of Eq. (\ref{re-mast}) and the coherent parametric amplification. The presence of the second term makes that the steady state is not a thermal equilibrium state anymore. We can prove that $\Delta X_{b}^{\theta2 }$ is minimized as $\Delta\hat{X}^2_b\simeq{\omega_b\langle\hat{n}_b\rangle_\textrm{ss} -\chi\over \omega_b'}+{1\over 2}$ when $\theta=0$ under the weak coupling condition. It is interesting to see that the squeezing is present as long as
\begin{equation}
\chi/\omega_b>\langle\hat{n}_b\rangle_\textrm{ss} ,  \label{squ-cri}
\end{equation}
which gives us a compact criteria for designing driving conditions to generate the stable phonon squeezing.

Via numerically solving Eq. (\ref{maseq}), we plot in Fig. \ref{squzphon}(a) the steady-state phonon number $\langle \hat{n}_b\rangle_\textrm{ss}$ as a function of $\Delta_m$ and $\Delta_c$. It is found that $\langle \hat{n}_b\rangle_\textrm{ss}$ is abruptly suppressed in the regime $\Delta_c'=\bar{\omega}$ [see the black dashed line in Fig. \ref{squzphon}(a)]. This is similar to the sideband cooling in cavity optomechanics \cite{PhysRevLett.99.093901,PhysRevLett.99.093902}. A more remarkable observation is that a large part of the parameter space supporting the decreased $\langle \hat{n}_b\rangle_\textrm{ss}$ is around $\Delta_m'=\bar{\omega}$ [see the white dot-dashed line in Fig. \ref{squzphon}(a)]. The condition $\Delta_c=0$ in this regime indicates that the phonon-number suppression here is dominated by the magnon-phonon interaction $G_{bm}$, which exhibits a dramatic photon-induced enhancement at $\Delta_c=0$, see Fig. \ref{fig1}(c). This result can be well explained by our analytical result in Eq. (\ref{phmb}), the denominator $\zeta_-^{r}-\zeta_+^{r}$ of whose leading term takes the maximum at $\Delta_{c/m}'=\bar{\omega}$.
Figure \ref{squzphon}(b) shows the fluctuations of $\hat{X}_b$ for different $\Omega_c$ and $\Delta_c$ via numerically solving Eq. (\ref{maseq}). It is interesting to find that the squeezing is present in the regimes with the suppressed phonon number at $\Delta^{\prime}_{c/m}=\bar{\omega}$. Its boundaries match exactly with our analytical criteria in the inequality (\ref{squ-cri}). The detailed comparison in Fig. \ref{squzphon}(c) further confirms that $\Delta X_b^2<1/2$ whenever $\chi/\omega_b>\langle\hat{n}_b\rangle_\textrm{ss}$, which validates our analytical criteria. Optimizing all possible $\Delta_{c/m}$, the minimal $\Delta X_b^2$ is decreased by increasing $\Omega_m$ [see Fig. \ref{squzphon}(d)], which efficiently enhances the parametric amplification interaction $\chi$. Thus our phonon squeezing can be further amplified by the magnetic driving. An important issue regarding the experimental observation of the squeezing is its robustness to the thermal fluctuation. Benefited from the suppression of phonon number, the optimal squeezing still persists even in the temperature order of $1$ K [see Fig. \ref{squzphon}(e)]. It is several times larger than the one in conventional cavity magnomechanics \cite{PhysRevA.99.021801}. This indicates the stability of the squeezing generated in our scheme. Therefore, we can manipulate the unique photon-phonon-magnon interactions to create a stable and robust phonon squeezing, which requires neither any extra nonlinearity nor the reservoir engineering \cite{PhysRevA.92.062311,PhysRevResearch.3.023126,PhysRevLett.127.083602,PhysRevA.99.021801}.

\begin{figure}[t]
	\centering
	\includegraphics[width=15cm]{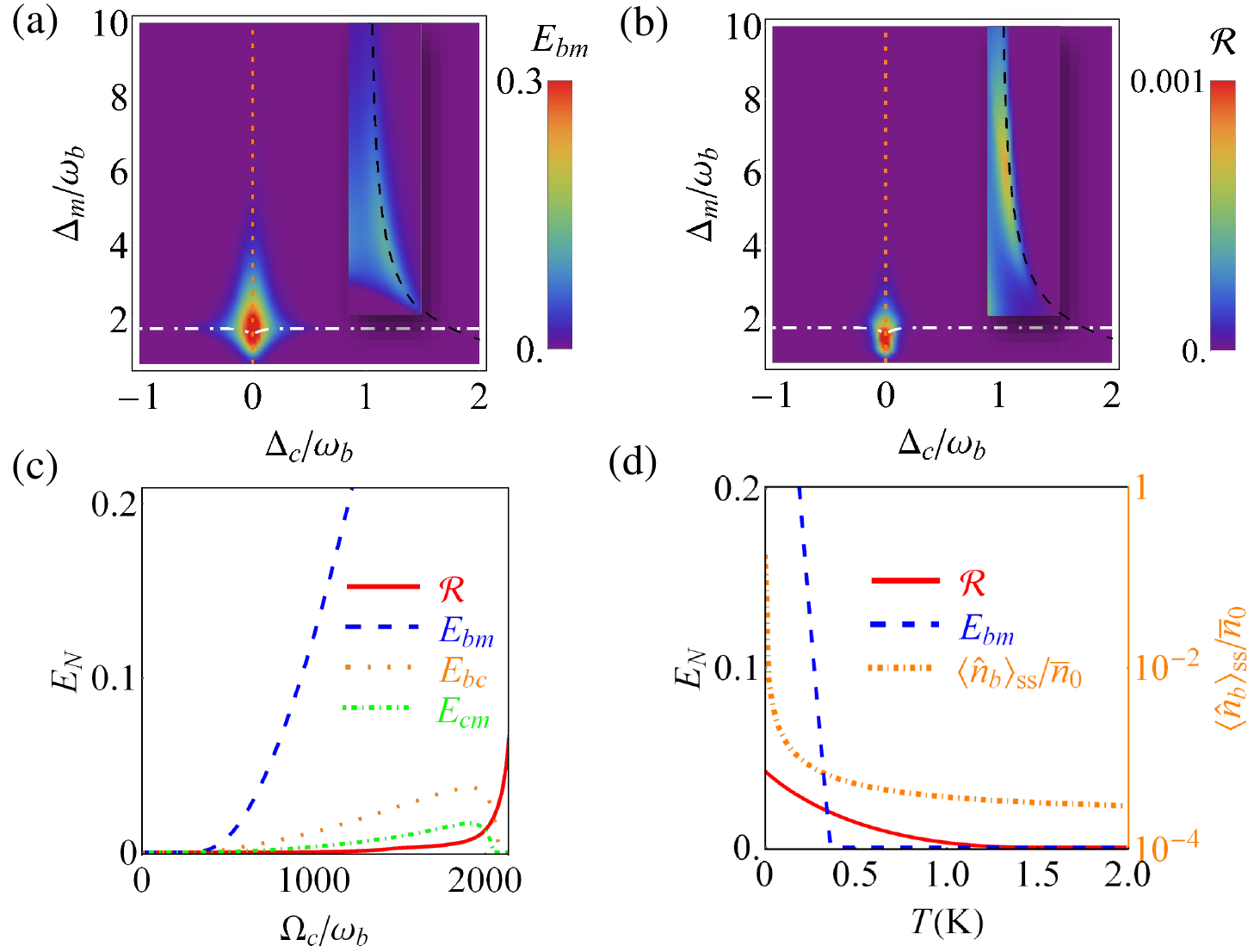}
	\caption{Magnon-phonon entanglement $E_{bm}$ (a) and tripartite entanglement $\mathcal{R}$ (b) as a function of $\Delta_m$ and $\Delta_c$ when $(\Omega_{c},\Omega_m)=(500,400)\omega_b$. The conditions of $\Delta_{c}=0$, $\Delta_{m}'=\bar\omega$, and $\Delta^{\prime}_{c}=\bar\omega$ are depicted by the orange dotted, white dot-dashed, and black dashed lines, respectively. The values in the boxes of (a) and (b) are magnified by $3000$ and $30000$ times. (c) Bi- and tri-partite entanglements as a function of $\Omega_c$ around the resonance with $(\Delta_{c},\Delta_{m})=(-0.28,1.4)\omega_b$ and $\Omega_m=400\omega_b$. (d) Temperature dependence of entanglements and $\langle \hat{n}_{b}\rangle_{\textrm{ss}}/\bar{n}_{0}$ when $\Omega_c=2100\omega_b$. Other parameters are the same as Figs. \ref{squzphon}.}	 \label{fig3}
\end{figure}

\section{Phonon-magnon-photon entanglement}\label{ent}
The coherent photon-phonon-magnon interactions also enable the generation of stable entanglement. We employ the logarithmic negativity \cite{PhysRevA.65.032314} and the minimum residual contangle \cite{Adesso_2007} to measure the bi- and tri-partite  entanglement, respectively. The logarithmic negativity between the $i$th and $j$th modes is defined as $E_{i|j}\equiv\max[0,-\log _{2}(2\tilde{\nu})]$, where $\tilde{\nu}=\min[\textrm{eig}(|i\Omega_{2}\tilde{\mathbf{V}}_{i|j}|)]$ is the minimum symplectic eigenvalue of the bipartite reduced covariance matrix $\tilde{\mathbf{V}}_{i|j}=\mathcal{P}_{i|j}\mathbf{V}_{i|j}\mathcal{P}_{i|j}$, $\Omega_2=\bigoplus_{j=1}^{2}i\sigma_y$ ($\sigma_y$ being the Pauli matrix) is the symplectic matrix, and $\mathcal{P}_{i|j}=\textrm{diag}(1,-1,1,1)$ is the partial transposition matrix. The minimum residual contangle is defined as $\mathcal{R}\equiv\min [\mathcal{R}^{i|jk},\mathcal{R}^{j|ik},\mathcal{R}^{k|ij}]$ with the residual contangle given by $\mathcal{R}^{i|jk}=E^2_{i|jk}-E^2_{i|j}-E^2_{i|k}$, where $E_{i|jk}$ is the one-mode-vs-two-mode logarithmic negativity.

The steady-state covariance matrix is obtained by solving Eq. (\ref{maseq}) (see \ref{con-mat}). Then the bi- and tri-partite entanglement are determined. Figures \ref{fig3}(a) and \ref{fig3}(b) show the magnon-phonon entanglement $E_{bm}$ and tripartite entanglement $\mathcal{R}$ in steady state as a function of $\Delta_m$ and $\Delta_c$, respectively. Being consistent to Fig. \ref{squzphon}, an obvious entanglement generation occurs in the regimes of $\Delta_{c}=0$ and $\Delta_{c/m}'=\bar\omega$. These entanglements are maximal in the resonant driving regime $\Delta_c=0$, where the magnon-phonon coupling is maximal [see Fig. \ref{fig1}(c)] and the phonon number is dramatically suppressed [see Fig. \ref{squzphon}(a)].
Because the effective magnon-phonon coupling strength $|G_{bm}|$ is much stronger than the photon-phonon one $|G_{bc}|$, the obtained $E_{bm}$ is larger than the photon-phonon entanglement $E_{bc}$ almost in one order of amplitude. We also find stable entanglements in the regime of $\Delta_c^{\prime}=\bar{\omega}$ (see the black dashed lines), however which is much smaller than the corresponding ones in the $\Delta_{c}=0$ regime. Figure \ref{fig3}(c) shows the optimized entanglements as a change of $\Omega_c$. We find that, with increasing $\Omega_c$, $E_{bm}$ and $\mathcal{R}$ monotonically increases, while other bi-partite entanglements decrease after increasing to the optimal values. The persistent enhancement of $E_{bm}$ and $\mathcal{R}$ is caused by the increasing of coupling strengths $|G_{bm}|$ and $|G_{bc}|$ with increasing $\Omega_c$. However, the phonon number is amplified with the increase of $\Omega_c$ over its minimum values, which results in the decreasing of other bipartite entanglements for large $\Omega_c$. The temperature dependence in Fig. \ref{fig3}(d) reveals that the entanglements survive up to about $1$ K for $\mathcal{R}$ and $0.4$ K for $E_{bm}$, which is one order larger than the one in conventional system \cite{PhysRevLett.121.203601}. Such enhancement is due to the nonlinear magnon-phonon interaction acting as an effective parametric amplification to the phonon \cite{PhysRevA.105.063704} and the accompanied suppression of phonon number [see Fig. \ref{fig3}(d)].

\section{Experimental realization}\label{dis}

Experimentally, optomechancial coupling in various low-dimensional systems have been realized, such as a graphene on a superconducting microwave cavity and a hexagonal boron nitride on a Si microdisk optical cavity \cite{PhysRevLett.113.027404,doi:10.1021/acs.nanolett.8b04956,Singh2014,Xie2021}. A new type of magnetically active mechanical oscillator has been recently built by 2D CrI$_{3}$ drumhead membranes \cite{Jiang2020,Pacchioni2020,Siskins2020}. Combing these experimental progresses, we expect a setup for the realization of our proposed cavity magnomechanics, by suspending a single-layer CrI$_{3}$ in the middle of an optical driven cavity. To clarify the physical relevance of the parameters used in calculations, we here provide a theoretical study on the phonon and magnon modes supported by a 2D magnet, as well as the magnon-phonon and photon-phonon coupling strengths with the practical material parameters.

In a rectangular geometry, the phonon modes of a membrane have analytical expressions with the resonance frequencies $\omega _{i,j}=\pi^{2} \sqrt{\epsilon/\rho}\sqrt{(i/L_{x})^{2}+(j/L_{y})^{2}}$ and wave vectors $\mathbf{k}=(i\pi/L_x,j\pi/L_y)$ \cite{HAUER2013181,mem}. Here, $\epsilon$ is the internal tensile stress, $\rho$ is the 2D mass density, and $L_{x}$ and $L_{y}$ are the lengths of membrane along the $x$- and $y$-directions. The typical parameters of the single-layer CrI$_{3}$ are $\sigma=2.04\times10^{18}$/m$^{2}$, $m_{0} =8.89\times 10^{-26}$ kg, $\rho=3.378\times 10^{-6}$ kg/m$^{2}$, $\epsilon = 0.53$ N/m, $S=3/2$, and $B_{1}=0.685$ meV \cite{zhang2015robust,PhysRevB.101.125111,doi:10.1126/science.aav4450,PhysRevB.101.125111}. To be specify, we chose $(L_{x},L_{y})=(0.4,0.17)$ mm, which leads to $\omega_b=25$ MHz for the $(1,1)$ mode with the wave vector $(k_x,k_y)=(7.8,18.5)$ km$^{-1}$. For the magnons, the dispersion relation of 2D ferromagnets on a honeycomb lattice is given by $\omega_{m}({\bf{q}})/\hbar=S[3J+2K-3J\sqrt{3+2\cos q_x+4\cos(q_{x}/2)\cos(\sqrt{3}q_{y}/2)}]+B$, where $J$ is the ferromagnetic Hersenberg exchange interaction, $K$ is the anisotropy, $B$ is the external magnetic field, and $\mathbf{q}=(q_{x},q_{y})$ is the wave vector. Typically, the values of decay rate for the phonons and magnons in magnet films at millikelvin temperature are on the orders to $10^{2}$ Hz and $10^{6}$ Hz \cite{Kosen2019,settipalli2023investigation}. In this manner, the amplitude of the driving magnetic field and the power of the input laser in obtaining the phonon squeezing and entanglement in Figs. \ref{squzphon} and \ref{fig3} range over $B_{0}\in (0.1,1)$ $\mu$T and $P\in(4,200)$ $\mu$W, which are experimentally available.

In structure with a suspended membrane in the middle of an optical driven cavity, we have the phonon-photon coupling strength $g_{bc}=\frac{\sin 2k_{j}x_{0}}{\sqrt{(1-T)^{-1}-\cos ^{2}2k_{j}x_{0}}}\frac{\omega _{j}}{L}\sqrt{\frac{\hbar }{2m_{b}\omega _{b}}}$ (see \ref{coupling-strength}), where $x_0$ is position of the membrane, $k_j=j\pi/L$ is wave vector of the $j$th photon mode with cavity length $L$, $T$ is the transparency, $m_b$ and $\omega_b$ are the effective mass and resonance frequency of the membrane. As can be seen, the transparency of the membrane has a significantly influence on the optomechanical couplings, which can be modified by coating opaque polymers on the membrane. Choosing $L=1$ mm, $T=0.95$ and the photon mode at wavelength $\lambda=1064$ nm, it leads to $g_{bc}= g_{bm}=10^{-5}\omega_b$, as used in Fig. \ref{squzphon}. In experiment, the decay rate for the photons is on the order to $10^{6}$ Hz \cite{Sankey2010}.

\section{Conclusion}\label{con}
In summary, we have proposed a realization of the cavity magnomechanics by a 2D magnetic material. The magnetoelastic interaction in the 2D material induces a unique {\color{red}{quadratic}} magnon-phonon coupling. Such nonlinear coupling contributes a parametric amplification and incoherent squeezing to the phonon, which makes our system having a natural superiority in quantum-state engineering. We find that a stable phonon squeezing, bi-partite and tri-partite entanglement are present in the resonant regime of optical driving, and the red-sideband regime of the optical and magnetic drivings. The accompanied suppression of phonon number endows these generated quantum features with the desired thermal stability. As the 2D material is compatible with the planar platform, our proposed system is helpful for the integration of 2D magnet as an important component in quantum optomechanical system, which may promote its applications in quantum operations and protocols. Enriching the physical platforms of cavity magnomechanics, our hybrid system may inspire the experimental exploration on macroscopic quantum magnomechanical effects in 2D materials.

\section{Data availability Statement}
All data that support the findings of this study are included within the article (and any supplementary files).

\ack
This work was supported by the National Natural Science Foundation of China (Grants  No. 12074106, No. 11904095, No. 12275109, No. 11834005, No. 12247101, and No. 12374178), the Science Fund for Distinguished Young Scholars of Hunan Province (Grant No. 2022JJ10002), the National Key Research and Development Program of Ministry of Science and Technology (2021YFA1200503), and the Fundamental Research Funds for the Central Universities from China.

\appendix

\section{Phonon-magnon interaction}\label{mag-pho-int}
The magnon-phonon interaction results from the magnetostrictive interaction. Specify, the deformation of magnetic material changes its magnetization, while the varying magnetization leads to the deformation of the magnetic material. Quite different from the one in sphere of yttrium iron garnet, the magnetostrictive coupling in the 2D magnetic membrane results in a quadratic magnon-phonon interacting Hamiltonian. To see it, we start from the magnetoelastic energy  \cite{PhysRevB.89.184413}
\begin{equation}
\mathcal{E}=\frac{\sigma }{\mathcal{M}^{2}_S}\int dV\sum_{\alpha\beta }B_{\alpha \beta }\mathcal{M}_{\alpha }\mathcal{M}_{\beta }U_{\alpha \beta }(\mathbf{r}), \label{en-1}
\end{equation}
where $\sigma=N/V$ is the number density of magnetic atoms, $\mathcal{M}_S$ is the saturation magnetization, $\mathcal{M}_{\alpha }$, with $\alpha=x,y,z$, are the local magnetization components, and $B_{\alpha \beta }$ are magnetoelastic coupling constants. The strain tensor of the thin film reads
\begin{equation}
U_{\alpha \beta }(\mathbf{r})=\frac{1}{2}[\partial _{\beta }U_{\alpha }(%
\mathbf{r})+\partial _{\alpha }U_{\beta }(\mathbf{r})+\sum_{\gamma }\partial
_{\alpha }U_{\gamma }(\mathbf{r})\partial _{\beta }U_{\gamma }(\mathbf{r})],
\label{smst}
\end{equation}
where $U_{\alpha }(\mathbf{r})$ are the components of the lattice displacement $\mathbf{U}(\mathbf{r})$. In 3D materials, the deformation is small and the second-order term in Eq. (\ref{smst}) are negligible \cite{cavitymagnonopto}. However, the second-order lattice deformation plays important roles in 2D materials because of its low mass, which has a large deformation under external perturbations \cite{PhysRevLett.129.123601}.

In two-dimensional lattice, neglecting the derivative of $z$, we have $\partial _{z}U_{z}=U_{zz}=0$. Considering the out-of-plane vibration mode $U_z(\mathbf{r})$, the interaction with the magnetization reads
\begin{eqnarray}
\mathcal{E} &=&\frac{\sigma B_{1}}{2\mathcal{M}_{S}^{2}}\int dxdy(\mathcal{M}_{x}^{2}+\mathcal{M}_{y}^{2})\partial _{x}^{2}U_{z}(\mathbf{r})
\nonumber \\
&&+\frac{\sigma B_{2}}{2\mathcal{M}_{S}^{2}}\int dxdy[(\mathcal{M}%
_{x}\mathcal{M}_{y}+\mathcal{M}_{y}\mathcal{M}_{x})\partial
_{x}U_{z}(\mathbf{r})\partial _{y}U_{z}(\mathbf{r})  \nonumber \\
&&+\sum_{\alpha =x,y}(\mathcal{M}_{\alpha }\mathcal{M}_{z}+\mathcal{M}_{z}%
\mathcal{M}_{\alpha })\partial _{\alpha }U_{z}(\mathbf{r})].\label{energy1}
\end{eqnarray}
where $B_{1}=B_{\alpha \alpha}$, $B_{2}=B_{\alpha \beta }$ ($\alpha\neq\beta$) and we have assumed $U_{xx}(\mathbf{r})=U_{yy}(\mathbf{r})$.
Via the Fourier transform $\mathbf{U}(\mathbf{r})=\int _{\mathbf{k}}e^{i\mathbf{k\cdot r}}\tilde{\mathbf{U}}(\mathbf{k})$, Eq. (\ref{energy1}) is recasted into
\begin{equation}
\mathcal{E}=\frac{\sigma }{2\mathcal{M}_{S}^{2}}\int_{\mathbf{k}%
}k_{x}[B_{1}k_{x}(\mathcal{M}_{x}^{2}+\mathcal{M}_{y}^{2})+B_{2}k_{y}(%
\mathcal{M}_{x}\mathcal{M}_{y}+\mathcal{M}_{y}\mathcal{M}_{x})]|\tilde{U}%
_{z}(\mathbf{k})|^{2},
\end{equation}
where $\int_{\mathbf{k}}=\int \frac{dk_{x}dk_{y}}{(2\pi )^{2}}$, $\mathbf{k}=(k_{x},k_{y})$ and $\mathbf{r}=(x,y)$. In derivation, we chose the spin wave in state with vector $\mathbf{q}=0$ and the momentum conservation relation of the displacement $\mathbf{k}+\mathbf{k}^{\prime }=0$  has been used. It implies that the annihilation of magnetic modes leads to the generation of two mechanical modes with wave vectors in opposite directions.

In order to make the quantization to the magnetic mode, we replace $\mathcal{M}_{\alpha}$ by the spin operator according to $\mathcal{M}_{\alpha}=2\mu_{B}\sigma \hat{S}_{\alpha}$, where $\mu_{B}$ is the Bohr magneton. Assuming the 2D magnet is magnetized with easy-axis anisotropy along the $z$ axis, the spin operators can be quantized by the Holstein-Primakoff transformation
\begin{equation}
\hat{S}_{x}\approx \frac{\sqrt{2S}}{2}(\hat{m}+\hat{m}^{\dag }),\hat{S}%
_{y}\approx \frac{\sqrt{2S}}{2i}(\hat{m}-\hat{m}^{\dag }),\hat{S}_{z}=S-\hat{%
m}^{\dag }\hat{m}.
\end{equation}
The quantized mode $\hat{m}$ is the magnon. Further introducing the creation and annihilation operators of the phonon mode $\tilde{U}_{z}(k)=\sqrt{\frac{\hbar }{2m_{0}\omega _{k}}}(\hat{b}_{k}+ \hat{b}_{k}^\dag)$, we obtain the quantized Hamiltonian
\begin{eqnarray}
\hat{H}_{bm} &=&\frac{B_{1}}{4S}\sum_{\mathbf{k}}\frac{\hbar k_{x}^{2}}{%
2m_{0}\omega _{k}}\hat{m}^{\dag }\hat{m}(\hat{b}_{k}+\hat{b}_{k}^{\dag })^{2}\nonumber
\\
&&-i\frac{B_{2}}{4S}\sum_{\mathbf{k}}\frac{\hbar k_{x}k_{y}}{2m_{0}\omega
_{k}}(\hat{m}^{2}-\hat{m}^{\dag 2})(\hat{b}_{k}+\hat{b}_{k}^{\dag })^{2},
\end{eqnarray}
where $m_{0}$ is the ion mass. In our study, the resonance frequency of the phonon is much smaller than that of the magnon. Thus, all the terms of $\hat{H}_{bm}$ in the second line are rapidly oscillating compared to the one in the first line in the interaction picture and are neglected under the rotating-wave approximation. We thus obtain the phonon-magnon interaction expressed in Eq. (\ref{hbm}).

\section{Phonon-photon interaction}\label{coupling-strength}
In configuration of two fixed mirrors at $\pm L$ and a movable 2D membrane positioned at $x_0$ with transmissivity $T$, the optomechanical coupling can be either linear or quadratic, depending on positions of the membrane \cite{PhysRevA.77.033819}. When the membrane is placed away from any nodes or antinodes of the cavity mode, i.e., $x_0\neq j\lambda_{j}/4$, the optomechanical interaction is governed by the traditional parametric coupling. In this case, the membrane divides the cavity into two sub-cavities. When $T\neq 0$, the coupling between the modes in the two-sub-cavities leads to the splitting of the initially degenerated optical modes $\omega_j=j\pi c/L$ into a pair of nondegenerated modes $\omega_{j,o/e}=\omega_{j}\pm\delta_{o/e}$, where the frequency shift is determined by $\delta _{e}=[\sin ^{-1}(\sqrt{1-T})-\sin ^{-1}(\sqrt{1-T}\cos 2k_{j}x_{0})]/\tau$ and $\delta _{o}=\pi /\tau -[\sin ^{-1}(\sqrt{1-T})+\sin ^{-1}(\sqrt{1-T}\cos2k_{j}q_{0})]/\tau $. $\tau=2L/c$ is the round trip time of photon in the cavity. Properly selecting the pumping frequency, we choose the even mode to be excited only. In this manner, we obtain the optomechanical interaction Hamiltonian $\hat{H}_{bc}=\hbar g_{bc}\hat{c}^{\dag}\hat{c}(\hat{b}^{\dag}+\hat{b})$, where $\hat{c}$ and $\hat{b}$ are the quantized photon and phonon mode. The coupling strength is determined by
\begin{equation}\label{gbc}
g_{bc}=\frac{\sin 2k_{j}x_{0}}{\sqrt{(1-T)^{-1}-\cos ^{2}2k_{j}x_{0}}}\frac{\omega _{j}}{L}\sqrt{\frac{\hbar }{2m_{b}\omega _{b}}},
\end{equation}
where $m_{b}$ and $\omega _{b}$ are the effective mass and resonance frequency of the mechanical mode.

\section{Reduced master equation} \label{red-mast}
Introducing the squeezing transformation $\hat{H}_{S}=\hat{S}^{\dag }\hat{H}\hat{S}$ with $\hat{S}=\exp[r(\hat{b}^{2}-\hat{b}^{\dag 2})/2]$ and $r=\textrm{arccoth}(\omega _{b}^{\prime }/2\chi)/2$, the linearized Hamiltonian can be reformed as $\hat{H}_{S}=\hat{H}_{0}+\hat{H}_{I}$ with $\hat{H}_{0}=\hbar\Delta _{c}^{\prime }\hat{c}^{\dag }\hat{c}+\hbar\Delta_{m}^{\prime}\hat{m}^{\dag }\hat{m}+\hbar\bar{\omega}\hat{b}^{\dag }\hat{b}$ and $\hat{H}_{I}=\hbar\sum_{x=c,m}e^{-r}(G_{bx}^{\ast }\hat{x}+G_{bx}\hat{x}^{\dag })(\hat{b}+\hat{b}^{\dag })$. Note that the phonon frequency has been modified by the nonlinear parametric amplification as $\bar\omega=\sqrt{\omega _{b}(\omega _{b}+4\chi )}$. Working in the interaction picture, the master equation of the total system is recast into
\begin{equation}
\dot{W}_{S}^{I}(t)=\frac{i}{\hbar}[W_{S}^{I}(t),\hat{H}_{I}(t)]+(\gamma _{c}\check{\mathcal{L}}_{c}+\gamma _{m}\check{\mathcal{L}}_{m}+\check{\mathcal{L}}_{S})W_{S}^{I}(t),  \label{A1}
\end{equation}
where $W_{S}^{I}(t)=e^{i\hat{H}_{0}t}\hat{S}^{\dag }W(t)\hat{S}e^{-i\hat{H}_{0}t}$ and $\hat{H}_{I}(t)=\hbar\sum_{x=c,m}\hat{A}_{x}^{\dag }(t)\hat{x}+\textrm{H.c.}$ with $\hat{A}_{x}(t)=e^{-r}G_{bx}e^{i\Delta ^{\prime} _{x}t}(\hat{b}e^{-i\bar{\omega}t}+\hat{b}^{\dag }e^{i\bar{\omega}t})$. The modified dissipator for the phonons is defined as $\check{\mathcal{L}}_{S}\cdot =\gamma _{b}[\bar{n}_{0}\cosh (2r)+\sinh
^{2}(r)](2\hat{b}^{\dag}\cdot\hat{b}-\hat{b}\hat{b}^{\dag}\cdot-\cdot\hat{b}\hat{b}^{\dag})+\gamma _{b}[\bar{n}_{0}\cosh(2r)+\cosh ^{2}(r)]2\hat{b}\cdot\hat{b}^{\dag}-\hat{b}^{\dag}\hat{b}\cdot-\cdot\hat{b}^{\dag}\hat{b}-[\gamma _{b}\sinh (r)\cosh(r)(2\bar{n}_{0}+1)e^{-2i\bar{\omega}t}(2\hat{b}\cdot\hat{b}-\hat{b}^{2}\cdot-\cdot\hat{b}^{2})+\textrm{H.c.}]$.

In the limit ($\gamma_m$, $\gamma_c\gg\gamma_b$) \cite{cavitymagnonopto,PhysRevLett.121.203601}, the photon and magnon modes rapidly decay to their ground state $(|0\rangle \langle 0|)_{c}\otimes(|0\rangle \langle 0|)_{m}$. We have assumed that the photon and magnon mode are un-correlated, because their coupling rate is rather small. In this manner, $W^{I}_{S}(t)$ can approximately factorize as $W_{S}^{I}(t)\simeq $\textrm{Tr}$_{c,m}[\rho _{S}(t)]\otimes (|0\rangle \langle 0|)_{c}(|0\rangle \langle 0|)_{m}$. Regarding the cavity and magnon modes as ``reservoir", one can adiabatically eliminate these two degrees of freedom based on the reservoir theory and obtain a reduced master equation satisfied by the mechanical oscillator. Explicitly, in the dissipation picture $\bar{W}_{S}^{I}(t)=e^{-(\check{\mathcal{L}}_{c}+\check{\mathcal{L}}_{m})t}W_{S}^{I}(t)$, Eq. (\ref{A1}) can be recast into
$\dot{\bar{W}}_{S}^{I}(t)=(\check{\mathcal{L}}_{I}(t)+\check{\mathcal{L}}_{S})\bar{W}_{S}^{I}(t)$, with $\check{\mathcal{L}}_{I}(t)\cdot =-\frac{i}{\hbar}e^{-(\check{\mathcal{L}}_{\hat{c}}+\check{\mathcal{L}}_{\hat{m}})t}[\hat{H}_{I}(t),\cdot ]e^{(\check{\mathcal{L}}_{\hat{c}}+\check{\mathcal{L}}_{\hat{m}})t}$. Under the Born-Markovian approximation, we obtain \cite{PhysRevA.92.062311}
\begin{eqnarray}
\dot{\bar{\rho}}_{S}^{I}(t) &=&\check{\mathcal{L}}_{S}\bar{\rho}_{S}^{I}(t)+\textrm{Tr}_{c,m}\int_{0}^{\infty }d\tau \check{\mathcal{L}}_{\textrm{I}}(t)%
\check{\mathcal{L}}_{\textrm{I}}(t-\tau )\nonumber \\&&\times \bar{\rho}_{S}^{I}(t)(|0%
\rangle \langle 0|)_{\textrm{c}}(|0\rangle \langle 0|)_{\textrm{m}},\label{A2}
\end{eqnarray}
where $\bar{\rho}_{S}^{I}(t)=\textrm{Tr}_{c,m}[\bar{W}_{S}^{I}(t)]$ and $\check{\mathcal{L}}_{I}(t)\cdot =-i\sum_{x=c,m}[\hat{\mathcal{A}}_{x+}(t)\hat{\mathcal{X}}_{-}(t)+\hat{\mathcal{A}}_{x-}(t)\hat{\mathcal{X}}_{+}(t)-\textrm{H.c.}]$, with $\hat{\mathcal{A}}_{x+}(t)=\hat{A}_{x}^{\dag }(t)\cdot ,\hat{\mathcal{A}}_{x-}(t)=\hat{A}_{x}(t)\cdot ,\hat{\mathcal{X}}_{+}(t)=e^{-\check{\mathcal{L}}_{\hat{x}}t}(\hat{x}^{\dag }\cdot )e^{\check{\mathcal{L}}_{\hat{x}}t}$, and $\hat{\mathcal{X}}_{-}(t)=e^{-\check{\mathcal{L}}_{\hat{x}}t}(\hat{x}\cdot )e^{\check{\mathcal{L}}_{\hat{x}}t}$. Making a time derivative to $\hat{\mathcal{X}}_{-}(t)$, we have $d\hat{\mathcal{X}}_{-}(t)/dt=-e^{-\check{\mathcal{L}}_{\hat{x}}t}[\check{\mathcal{L}}_{\hat{x}},\hat{x}\cdot ]e^{\check{\mathcal{L}}_{\hat{x}}t}$. One can easily check $[\check{\mathcal{L}}_{\hat{x}},\hat{x}\cdot ]=\gamma _{x}\hat{x}\cdot $, we then obtain
\begin{equation}
\hat{\mathcal{X}}_{-}(t)=e^{-\gamma _{x}t}(\hat{x}\cdot ), \label{A3}
\end{equation}
which also gives
\begin{equation}
\hat{\mathcal{X}}_{-}^{\dag }(t)=e^{-\gamma _{x}t}(\cdot \hat{x}^{\dag }). \label{A4}
\end{equation}
With the similar manner, we have $d\hat{\mathcal{X}}_{+}(t)/dt=-e^{-\check{\mathcal{L}}_{\hat{x}}t}[\check{\mathcal{L}}_{x},\hat{x}^{\dag }\cdot ]e^{-\check{\mathcal{L}}_{\hat{x}}t}$. From the commutation relation $[\check{\mathcal{L}}_{x},\hat{x}^{\dag }\cdot ]=2\gamma _{x}\cdot \hat{x}^{\dag }-\gamma _{x}\hat{x}^{\dag }\cdot $, it can be recast into $d\hat{\mathcal{X}}_{+}(t)/dt=-2\gamma _{x}\hat{\mathcal{X}}_{-}^{\dag}(t)+\gamma _{x}\hat{\mathcal{X}}_{+}(t)$. In the form of Eq. (\ref{A4}), we obtain
\begin{equation}
\hat{\mathcal{X}}_{+}(t)=e^{\gamma _{x}t}(\hat{x}^{\dag }\cdot )+(e^{-\gamma
_{x}t}-e^{\gamma _{x}t})(\cdot \hat{x}^{\dag }), \label{A5}
\end{equation}
which also results
\begin{equation}
\hat{\mathcal{X}}_{+}^{\dag }(t)=e^{\gamma _{x}t}(\cdot \hat{x})+(e^{-\gamma
_{x}t}-e^{\gamma _{x}t})(\hat{x}\cdot ). \label{A6}
\end{equation}

From the obtained forms of Eqs. (\ref{A3}), (\ref{A4}), (\ref{A5}), (\ref{A6}),
we have the nonzero correlation functions of the cavity and magnon field
\begin{eqnarray}
\langle \hat{\mathcal{X}}_{-}(t)\hat{\mathcal{X}}_{+}(t-\tau )\rangle
&=&\langle \hat{\mathcal{X}}_{+}(t)\hat{\mathcal{X}}_{+}^{\dag }(t-\tau
)\rangle =e^{-\gamma _{x}\tau },\label{C7} \\
\langle \hat{\mathcal{X}}_{-}^{\dag }(t)\hat{\mathcal{X}}_{+}^{\dag }(t-\tau
)\rangle  &=&\langle \hat{\mathcal{X}}_{+}^{\dag }(t)\hat{\mathcal{X}}%
_{+}(t-\tau )\rangle =e^{-\gamma _{x}\tau },\label{C8}
\end{eqnarray}
where $\langle \cdot \rangle =\textrm{Tr}_{c,m}[\cdot (|0\rangle \langle 0|)_{c}(|0\rangle
\langle 0|)_{m}]$.

Substituting Eq. (\ref{C7}) and Eq. (\ref{C8}) into Eq. (\ref{A2}), we obtain
\begin{eqnarray}
&&\textrm{Tr}_{c,m}\int_{0}^{\infty }\check{\mathcal{L}}_{I}(t)\check{\mathcal{L}}_{I}(t-\tau )\bar{\rho}_{S}^{I}(t)(|0\rangle \langle 0|)_{c}(|0\rangle\langle 0|)_{\textrm{m}}d\tau  \nonumber \\
&=&\int_{0}^{\infty }d\tau \sum_{x=c,m}e^{-\gamma _{x}\tau }[\hat{A}_{x}(t)\bar{\rho}_{S}^{I}(t)\hat{A}_{x}^{\dag }(t-\tau ) \nonumber\\
&&-\hat{A}_{x}^{\dag }(t)\hat{A}_{x}(t-\tau )\bar{\rho}_{S}^{I}(t)+\textrm{H.c.}].   \label{A8}
\end{eqnarray}
Remembering the form of $\hat{A}_{x}$ and returning back to the Schr\"{o}dinger picture, we have
\begin{eqnarray}
&&\int_{0}^{\infty }e^{-\gamma _{x}\tau }e^{-i\hat{H}_{0}t}\hat{A}_{x}(t)%
\bar{\rho}_{S}^{I}(t)\hat{A}_{x}^{\dag }(t-\tau )e^{i\hat{H}_{0}t}d\tau  \nonumber\\
&=&|G_{bx}e^{-r}|^{2}(\hat{b}+\hat{b}^{\dag })\rho _{S}(t)(h_{x+}\hat{b}%
+h_{x-}\hat{b}^{\dag }),\label{C10}
\end{eqnarray}
with $h_{x\pm }=\frac{1}{\gamma _{x}-i(\Delta^{\prime} _{x}\pm \bar{\omega})}$. In obtaining Eq. (\ref{C10}), the integral identity $\int_{0}^{\infty }d\tau e^{-[\gamma _{x}-i(\Delta^{\prime} _{x}\pm \bar{\omega})]\tau }=\frac{1}{\gamma _{x}-i(\Delta^{\prime} _{x}\pm \bar{\omega})}$ have been used. The other terms in Eq. (\ref{A8}) can be calculated in the similar manner. Making the inverse squeezing transformation and reback to the Schr\"{o}dinger picture, we finally arrive at the reduced master equation used in the maintext.

\section{Covariance matrix}\label{con-mat}
The covariance matrix of the photon-phonon-magnon system is defined as
\begin{equation}
V_{ij}=\langle \Delta \hat{\mathcal{F}}_{i}\Delta \hat{\mathcal{F}}_{j}+\Delta \hat{\mathcal{F}}_{j}\Delta \hat{\mathcal{F}}_{i}\rangle /2, \label{con}
\end{equation}
where the fluctuation operators $\Delta \hat{\mathcal{F}}_{i}=\hat{\mathcal{F}}_{i}-\langle \hat{\mathcal{F}}_{i}\rangle $ are determined by the matrix $\hat{\mathcal{F}}=(\hat{X}_{b},\hat{Y}_{b},\hat{X}_{m},\hat{Y}_{m},\hat{X}_{c},\hat{Y}_{c})$ with the quadrature operators $\hat{X}_{m}=(\hat{m}+\hat{m}^{\dag })/\sqrt{2}$, $\hat{Y}_{m}=(\hat{m}-\hat{m}^{\dag })/\sqrt{2}i$, $\hat{X}_{c}=(\hat{c}+\hat{c}^{\dag })/\sqrt{2}$, and $\hat{Y}_{c}=(\hat{c}-\hat{c}^{\dag })/\sqrt{2}i$. For the bipartite system, its covariance matrix can be readily obtained by straightforwardly eliminating the relevant rows and columns in Eq. (\ref{con}).

From the master equation, the equations of motion of the elements are determined by
\begin{eqnarray}
\fl\partial _{t}V_{11} &=&-2\gamma _{b}V_{11}-2p_{-}V_{12}+\gamma
_{b}+2\gamma _{b}\bar{n}_{0}+1/2  \nonumber \\
\fl\partial _{t}V_{22} &=&-2p_{+}V_{12}-2\gamma
_{b}V_{22}-4[G_{mb}^{r}V_{23}+G_{mb}^{i}V_{24}+G_{bc}^{r}V_{25}+G_{bc}^{i}V_{26}]+\gamma _{b}(2%
\bar{n}_{0}+1)+1/2  \nonumber \\
\fl\partial _{t}V_{33} &=&4G_{mb}^{i}V_{13}-2\gamma _{m}V_{33}+2\Delta
_{m}^{\prime }V_{34}+\gamma _{m}+1/2  \nonumber \\
\fl\partial _{t}V_{44} &=&-4G_{mb}^{r}V_{14}-2\Delta _{m}^{\prime
}V_{34}-2\gamma _{m}V_{44}+\gamma _{m}+1/2  \nonumber \\
\fl\partial _{t}V_{55} &=&4G_{bc}^{i}V_{15}-2\gamma _{c}V_{55}+2\Delta
_{c}^{\prime }V_{56}+\gamma _{c}+1/2  \nonumber \\
\fl\partial _{t}V_{66} &=&-4G_{bc}^{r}V_{16}-2\gamma _{c}V_{66}-2\Delta
_{c}^{\prime }V_{56}+\gamma _{c}+1/2  \nonumber \\
\fl\partial _{t}V_{12} &=&-p_{+}V_{11}-p_{-}V_{22}-2[\gamma
_{b}V_{12}+G_{mb}^{r}V_{13}+G_{mb}^{i}V_{14}+G_{bc}^{r}V_{15}+G_{bc}^{i}V_{16}]
\nonumber \\
\fl\partial _{t}V_{13} &=&2G_{mb}^{i}V_{11}-\gamma _{bm}V_{13}+\Delta
_{m}^{\prime }V_{14}-p_{-}V_{23}  \nonumber \\
\fl\partial _{t}V_{14} &=&-2G_{mb}^{r}V_{11}-\Delta _{m}^{\prime
}V_{13}-\gamma _{bm}V_{14}-p_{-}V_{24}  \nonumber \\
\fl\partial _{t}V_{15} &=&2G_{bc}^{i}V_{11}-\gamma _{bc}V_{15}+\Delta
_{c}^{\prime }V_{16}-p_{-}V_{25}  \nonumber \\
\fl\partial _{t}V_{16} &=&-2G_{bc}^{r}V_{11}-\Delta _{c}^{\prime
}V_{15}-\gamma _{bc}V_{16}-p_{-}V_{26}  \nonumber \\
\fl\partial _{t}V_{23}
&=&-2[G_{mb}^{r}V_{33}-G_{mb}^{i}V_{12}+G_{mb}^{i}V_{34}+G_{bc}^{r}V_{35}+G_{bc}^{i}V_{36}]-p_{+}V_{13}-\gamma _{bm}V_{23}+\Delta _{m}^{\prime }V_{24}
\nonumber \\
\fl\partial _{t}V_{24}
&=&-2[G_{mb}^{i}V_{44}+G_{mb}^{r}V_{12}+G_{mb}^{r}V_{34}+G_{bc}^{r}V_{45}+G_{bc}^{i}V_{46}]-p_{+}V_{14}-\gamma _{bm}V_{24}-\Delta _{m}^{\prime }V_{23}
\nonumber \\
\fl\partial _{t}V_{25}
&=&-2[G_{bc}^{r}V_{55}-G_{bc}^{i}V_{12}+G_{mb}^{r}V_{35}+G_{mb}^{i}V_{45}+G_{bc}^{i}V_{56}]-p_{+}V_{15}-\gamma _{bc}V_{25}+\Delta _{c}^{\prime }V_{26}
\nonumber \\
\fl\partial _{t}V_{26}
&=&-2[G_{bc}^{i}V_{66}+G_{bc}^{r}V_{12}+G_{mb}^{r}V_{36}+G_{mb}^{i}V_{46}+G_{bc}^{r}V_{56}]-p_{+}V_{16}-\gamma _{bc}V_{26}-\Delta _{c}^{\prime }V_{25}
\nonumber \\
\fl\partial _{t}V_{34} &=&-\Delta _{m}^{\prime }V_{33}+\Delta _{m}^{\prime
}V_{44}-2G_{mb}^{r}V_{13}+2G_{mb}^{i}V_{14}-2\gamma _{m}V_{34}  \nonumber \\
\fl\partial _{t}V_{35} &=&2G_{bc}^{i}V_{13}+2G_{mb}^{i}V_{15}-\gamma
_{mc}V_{35}+\Delta _{c}^{\prime }V_{36}+\Delta _{m}^{\prime }V_{45}
\nonumber \\
\fl\partial _{t}V_{36} &=&-2G_{bc}^{r}V_{13}+2G_{mb}^{i}V_{16}-\Delta
_{c}^{\prime }V_{35}-\gamma _{mc}V_{36}+\Delta _{m}^{\prime }V_{46}
\nonumber \\
\fl\partial _{t}V_{45} &=&2G_{bc}^{i}V_{14}-2G_{mb}^{r}V_{15}-\Delta
_{m}^{\prime }V_{35}-\gamma _{mc}V_{45}+\Delta _{c}^{\prime }V_{46}
\nonumber \\
\fl\partial _{t}V_{46} &=&-2G_{bc}^{r}V_{14}-2G_{mb}^{r}V_{16}-\Delta
_{m}^{\prime }V_{36}-\Delta _{c}^{\prime }V_{45}-\gamma _{mc}V_{46}
\nonumber \\
\fl\partial _{t}V_{56} &=&-\Delta _{c}^{\prime }V_{55}+\Delta _{c}^{\prime
}V_{66}-2G_{bc}^{r}V_{15}+2G_{bc}^{i}V_{16}-2\gamma _{c}V_{56}.
\end{eqnarray}
where $P_{\pm }=2\chi \pm \omega _{b}^{\prime },\gamma _{bm}=\gamma_{b}+\gamma _{m}$, $\gamma _{bc}=\gamma _{b}+\gamma _{c}$, and $\gamma _{mc}=\gamma _{m}+\gamma _{c}$. Then the covariance matrix can be numerically calculated. Note that, to ensure the stability of the system, the Routh-Hurwitz criteria has to be satisfied in the calculation \cite{PhysRevLett.98.030405}.

\section*{References}
\bibliographystyle{iopart-num}
\bibliography{mg}

\end{document}